\documentclass[prb,twocolumn,preprintnumbers,amsmath,amssymb,superscriptaddress]{revtex4}

\usepackage{graphicx}
\usepackage{bm}

\begin{document}

\title{Computation of dynamical correlation functions of Heisenberg chains in a field}

\author{Jean-S\'ebastien Caux} 
\affiliation{Institute for Theoretical Physics, University of
  Amsterdam, 
  1018 XE Amsterdam, The Netherlands.}
\author{Jean-Michel Maillet} 
\affiliation{Laboratoire de Physique, \'Ecole Normale Sup\'erieure de Lyon, 
  69364 Lyon, France.}

\date{\today}

\begin{abstract}
We compute the momentum- and frequency-dependent longitudinal spin structure factor for the 
one-dimensional spin-1/2 $XXZ$ Heisenberg spin chain in a magnetic field, using exact determinant
representations for form factors on the lattice.  
Multiparticle contributions are computed numerically throughout the Brillouin zone, yielding
saturation of the sum rule to high precision.  
\end{abstract}

\maketitle

Exact solutions of quantum models, either on the lattice or in the continuum, are
invaluable in the study of nonperturbative aspects of low-dimensional physics.
Bethe's construction of the full set of eigenstates of Heisenberg's spin exchange model \cite{BetheZP71}, 
by the method today commonly known as the Bethe Ansatz \cite{KorepinBOOK}, has 
ultimately led to a wide variety of exact results on the thermodynamics of integrable models,
with a broad spectrum of applications\cite{MattisBOOK,EsslerREVIEW}.

A long-standing limitation of the Bethe Ansatz was that the dynamics of such models was not 
directly accessible.  In recent years, however, enormous progress has been made 
for correlation functions of the Heisenberg spin chain \cite{JimboBOOK,MailletNPB554,MailletNPB567}.
In particular, matrix elements of any
local operator between two Bethe states can now be written as matrix determinants \cite{MailletNPB554}.  
Combined with formulas for
eigenstate norms \cite{KorepinCMP86}, this yields exact expressions for form factors on the lattice,
thereby opening the door to the computation of dynamical correlation functions.

In this paper, we implement this program for one of the cornerstones of the theory of integrable models, 
the anisotropic Heisenberg antiferromagnetic chain in a magnetic field:
\begin{eqnarray}
H \!=\! J \!\sum_{j = 1}^N \!\left[ S^x_j S^x_{j+1} \!+\! S^y_j S^y_{j+1} \!+\! \Delta \!\left(S^z_j S^z_{j+1}\!-\!\frac{1}{4}\right) \!-\! h S^z_j \right]
\label{XXZ}
\end{eqnarray}
with periodic boundary conditions.  The anisotropy parameter $\Delta$ will be chosen to lie in
the gapless regime $-1 < \Delta \leq 1$, and therefore the model provides a well-controlled 
realization of quantum critical behaviour.

Our interest lies in the space- and time-dependent spin 
correlation functions.  We will numerically compute 
the longitudinal dynamical spin structure factor, which is defined as the Fourier transform of the
spin-spin correlation function:
\begin{eqnarray}
S^{zz} (q, \omega) = \!\frac{1}{N} \!\sum_{j, j'=1}^N \!e^{i q(j - j')} \!\int_{-\infty}^{\infty} dt e^{i\omega t} 
\langle S^z_j(t) S^z_{j'}(0) \rangle_c
\label{S_zz}
\end{eqnarray}
where the subscript $c$ means that we take the connected part.  
This quantity is directly accessible experimentally through neutron scattering (see {\it e.g.}
[\onlinecite{KenzelmannPRB65},\onlinecite{StonePRL91}]).  

The longitudinal and transverse correlations for the isotropic Heisenberg chain in a field were studied
at fixed momentum $q$ in [\onlinecite{BiegelEPL59}], and the transverse ones for $XXZ$ at zero field and at $q = \pi$ in [\onlinecite{BiegelJPA36}].
More recently, the two-particle contributions to the longitudinal structure factor for the $XXZ$ chain in a field at
$q = \pi/2$ were studied for all energies in [\onlinecite{SatoJPSJ73}].  In particular, this 
allowed (at low energies) for a numerical check of 
conformal scaling exponents \cite{KorepinBOOK} to an impressive degree of precision.  
Here, we present results for all momenta $q$, including multiparticle contributions.  This yields data beyond the reach of 
conformal field theory, and allows us to quantitatively evaluate the precision of our results via the sum rule.

The exact solution through the Bethe Ansatz for model (\ref{XXZ}) is well-known \cite{KorepinBOOK}.  
The reference state is taken to be the state with all spins up, 
$|0\rangle = \otimes_{i = 1}^N |\uparrow\rangle_i$.  The Hilbert space separates into subspaces of fixed magnetization,
determined from the number of reversed spins $M$.  We take the number of sites $N$ and of reversed spins
$M$ to both be even.  Eigenstates in each subspace are completely determined for $2M \leq N$ by a set of rapidities $\{\lambda_j\}$, 
$j = 1, ...,M$, solution to the Bethe equations
\begin{eqnarray}
\mbox{atan} \!\!\left[ \frac{\tanh(\lambda)}{\tan(\zeta/2)} \right] \!\!-\! \frac{1}{N} \!\sum_{k = 1}^M 
\mbox{atan} \!\!\left[\frac{\tanh(\lambda - \lambda_k)}{\tan \zeta}\right] \!\!=\! \pi \frac{I_j}{N}
\label{BAE_log}
\end{eqnarray}
where $\Delta = \cos \zeta$.  Each choice of a set of distinct half-integers $\{I_j\}$, $j = 1, ...,M$ (with $I_j$ defined mod$(N)$)
uniquely specifies a set of rapidities, and therefore an eigenstate.  The energy and momentum of a state are
\begin{eqnarray}
E &=& J \sum_{i = 1}^M \frac{-\sin^2 \zeta}{\cosh 2\lambda_i - \cos \zeta} - h(\frac{N}{2} - M), \nonumber \\
q &=& \pi M - \frac{2\pi}{N}\sum_{i = 1}^M I_i \hspace{0.5cm} \mbox{mod} \hspace{0.2cm}2\pi.
\label{Ep}
\end{eqnarray}
The ground state is given by $I_j^0 = -\frac{M+1}{2} + j$, $j = 1, ...,M$. 

In terms of form factors, the structure factor can be written as a sum
\begin{eqnarray}
S^{zz} (q, \omega) = 2\pi \sum_{\alpha \neq 0} |\langle 0 | S_q^z | \alpha \rangle |^2 \delta(\omega -\omega_{\alpha})
\label{S_zz_ff}
\end{eqnarray}
over the whole set of 
eigenstates (distinct from the ground state) in the fixed $M$ subspace.  
Each term in (\ref{S_zz_ff}) can be obtained \cite{MailletNPB554} as a product of determinants of 
$M$-dimensional matrices, fully determined for a given eigenstate by a knowledge of the corresponding set of rapidities.
For the sake of brevity we do not reproduce the expressions for these matrices here.  

\begin{figure*}
\begin{tabular}{ll}
\includegraphics[width=8cm]{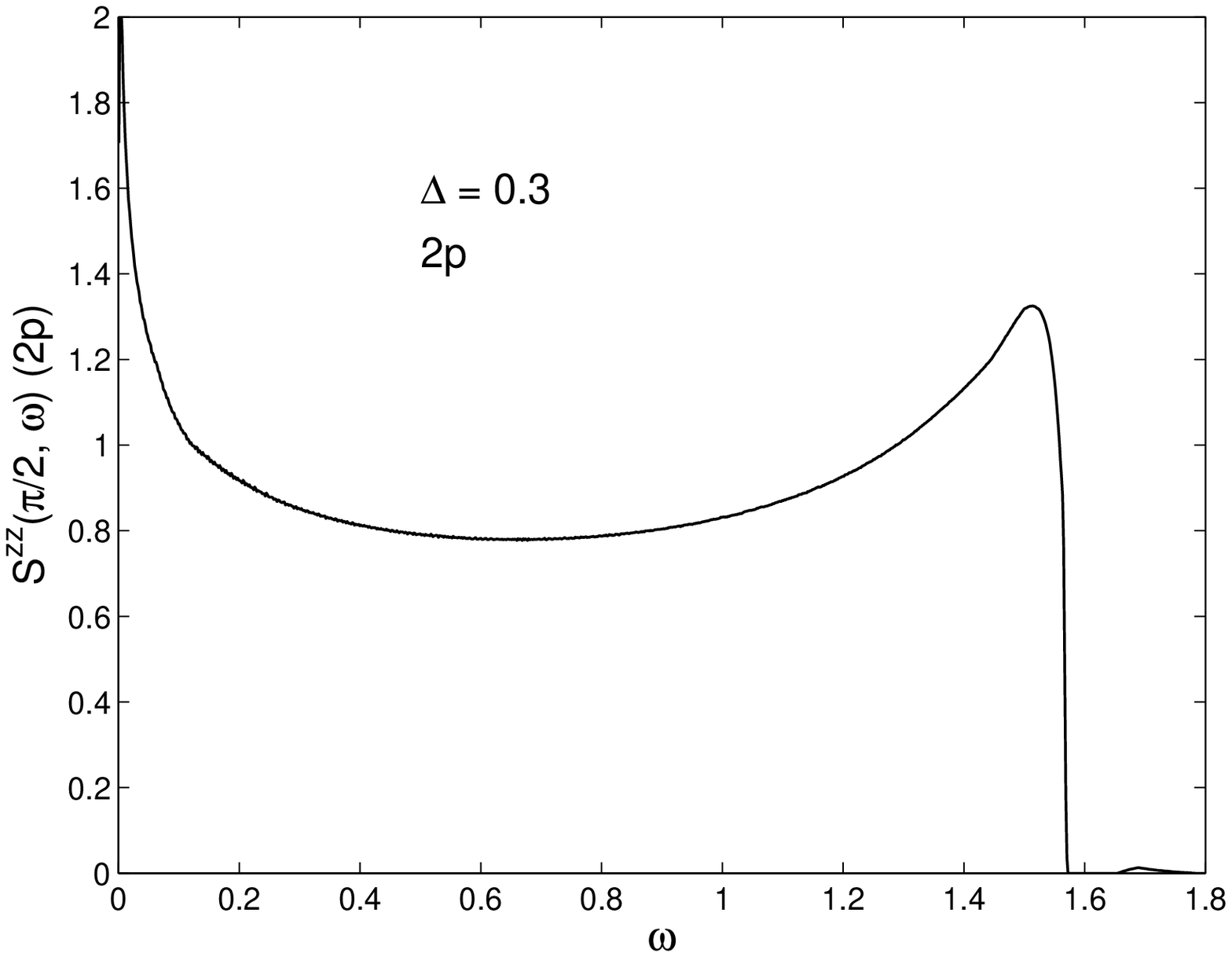}
& \includegraphics[width=8cm]{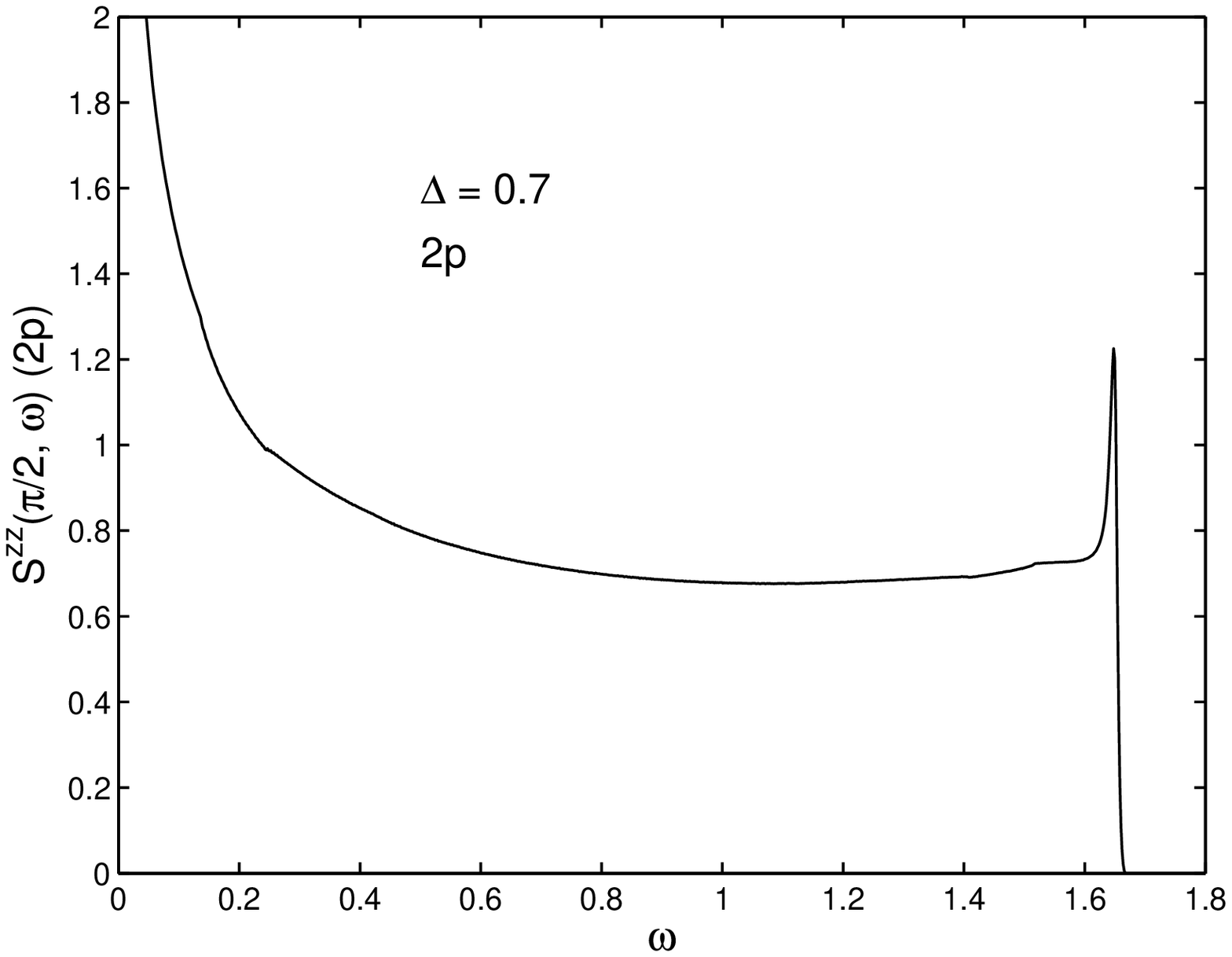} \\
\includegraphics[width=8cm]{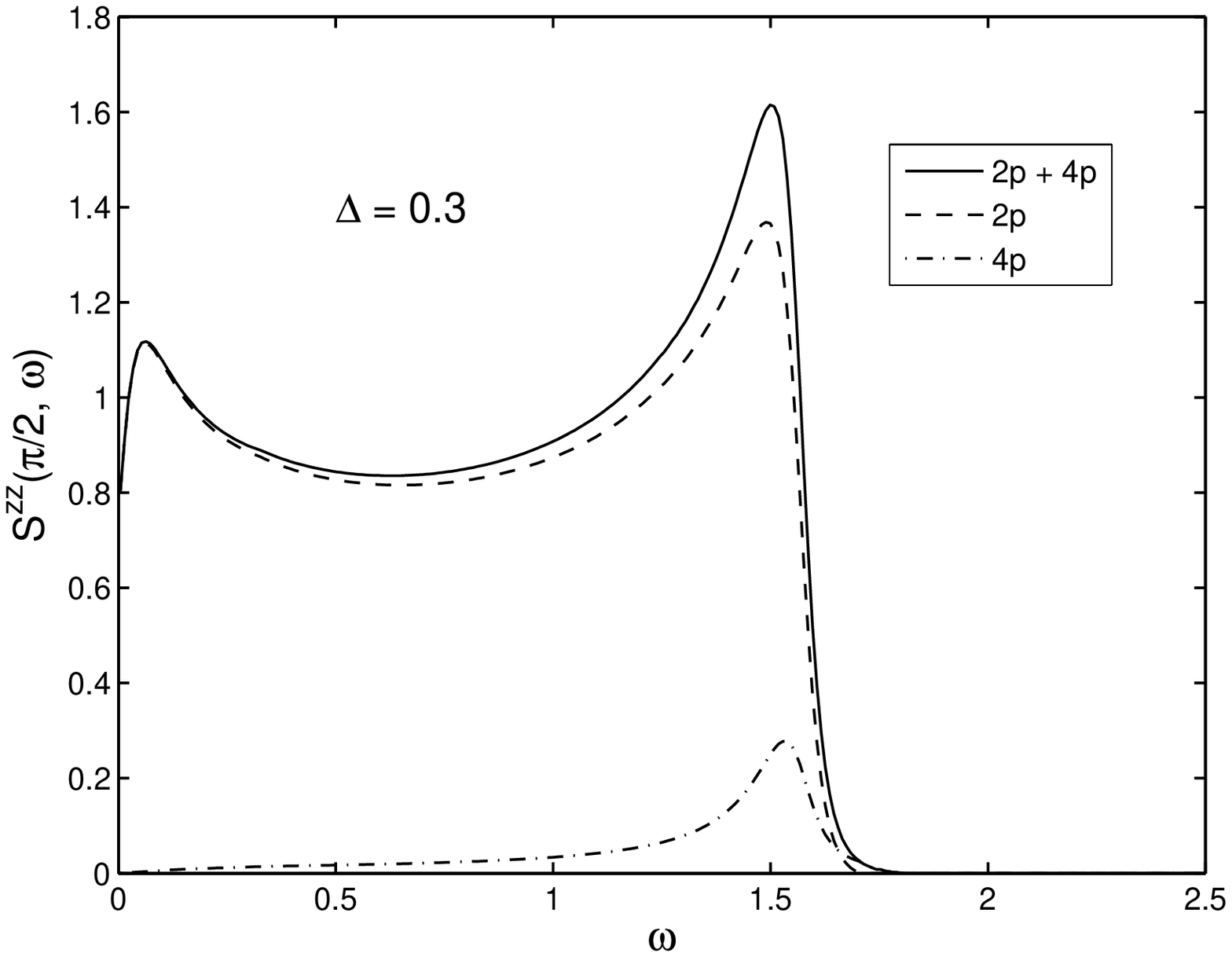}
& \includegraphics[width=8cm]{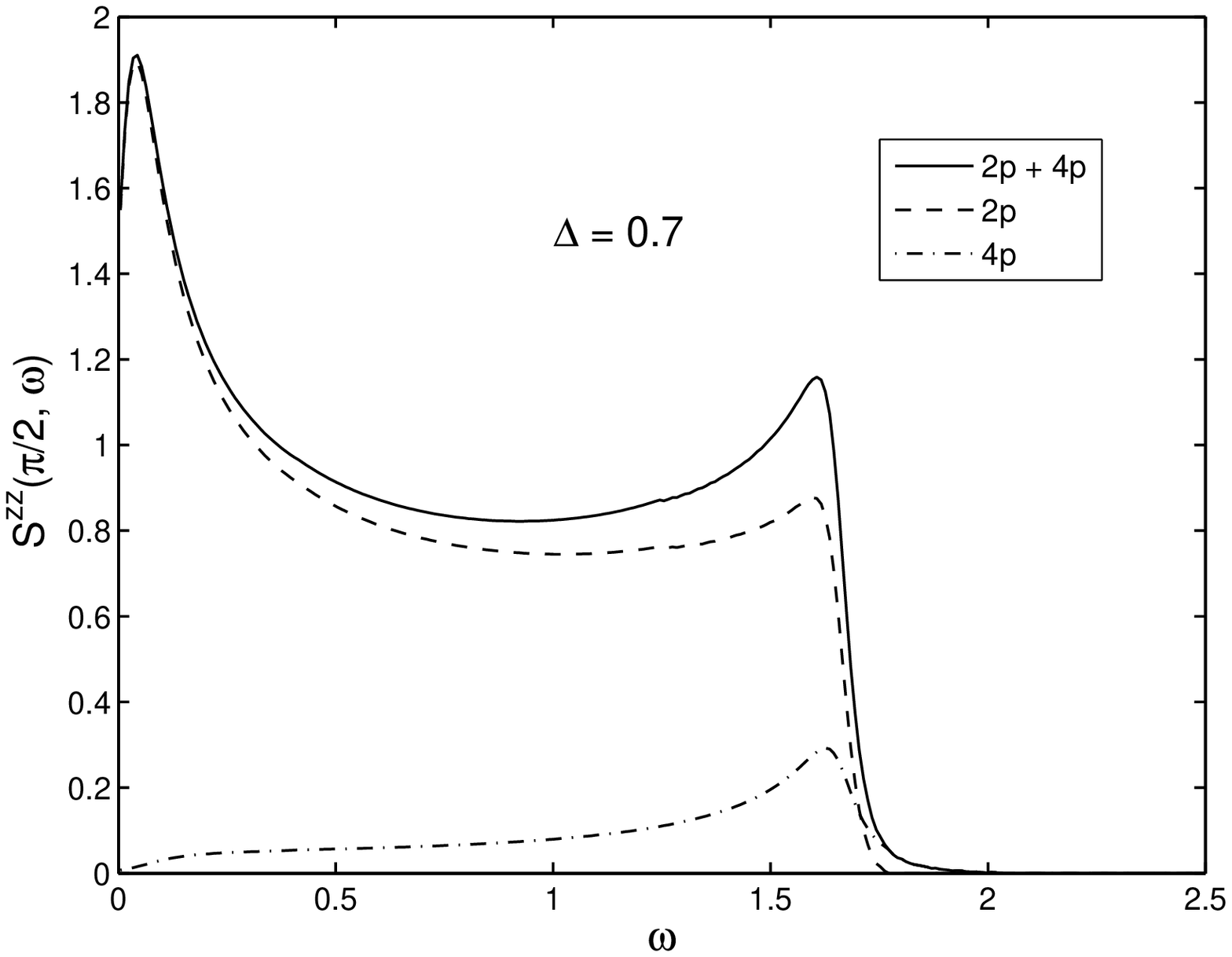} \\
\includegraphics[width=8cm]{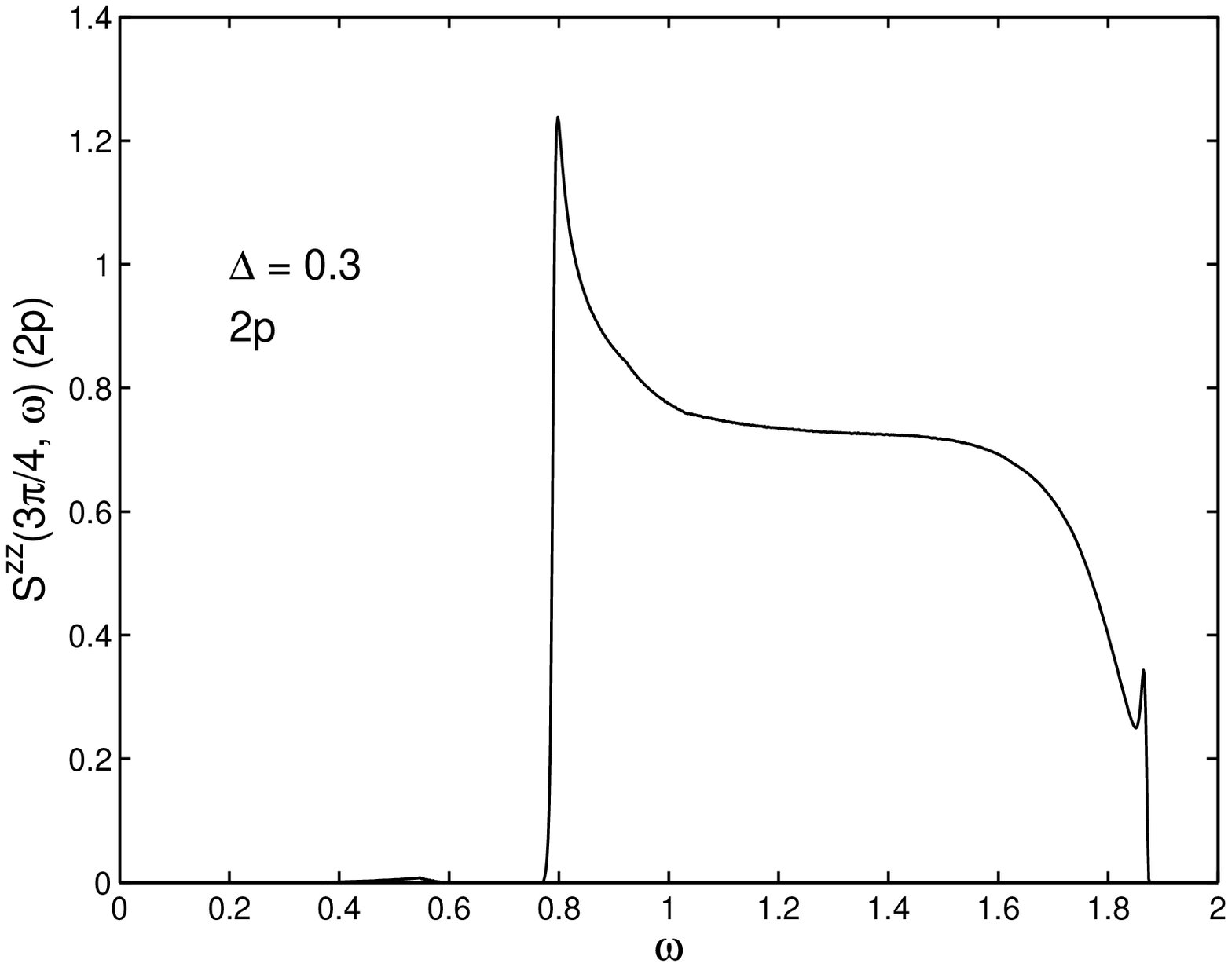}
& \includegraphics[width=8cm]{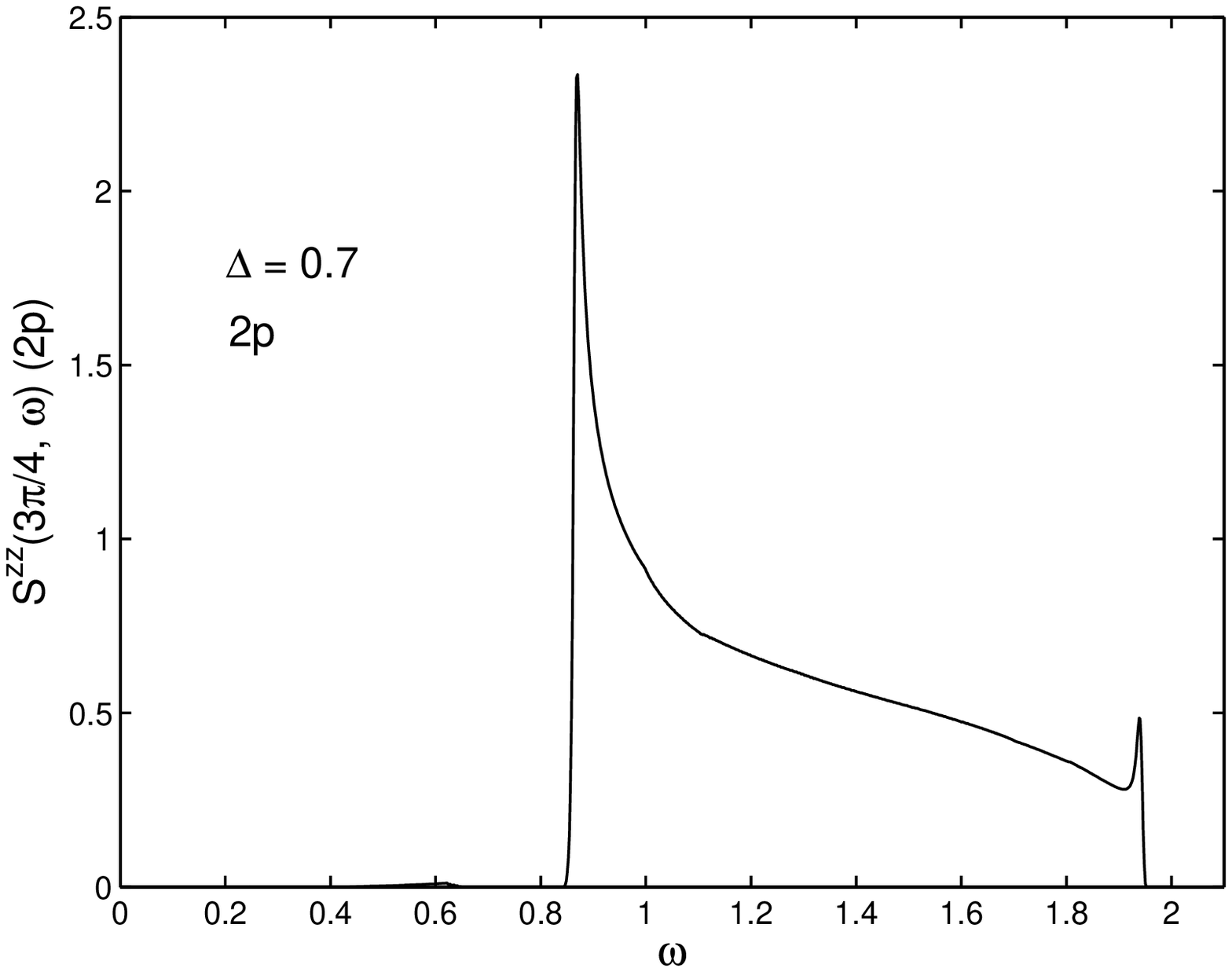} \\
\includegraphics[width=8cm]{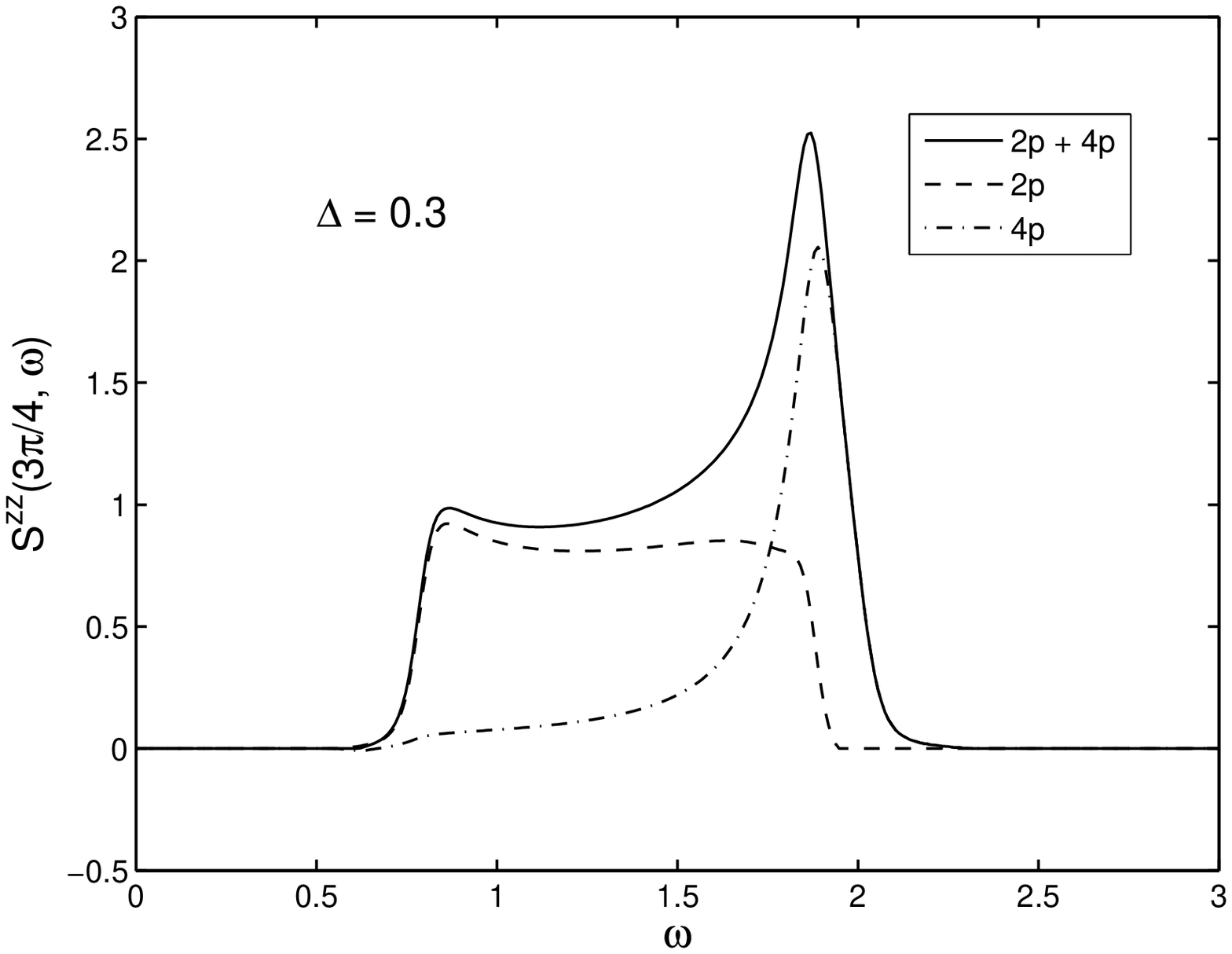}
& \includegraphics[width=8cm]{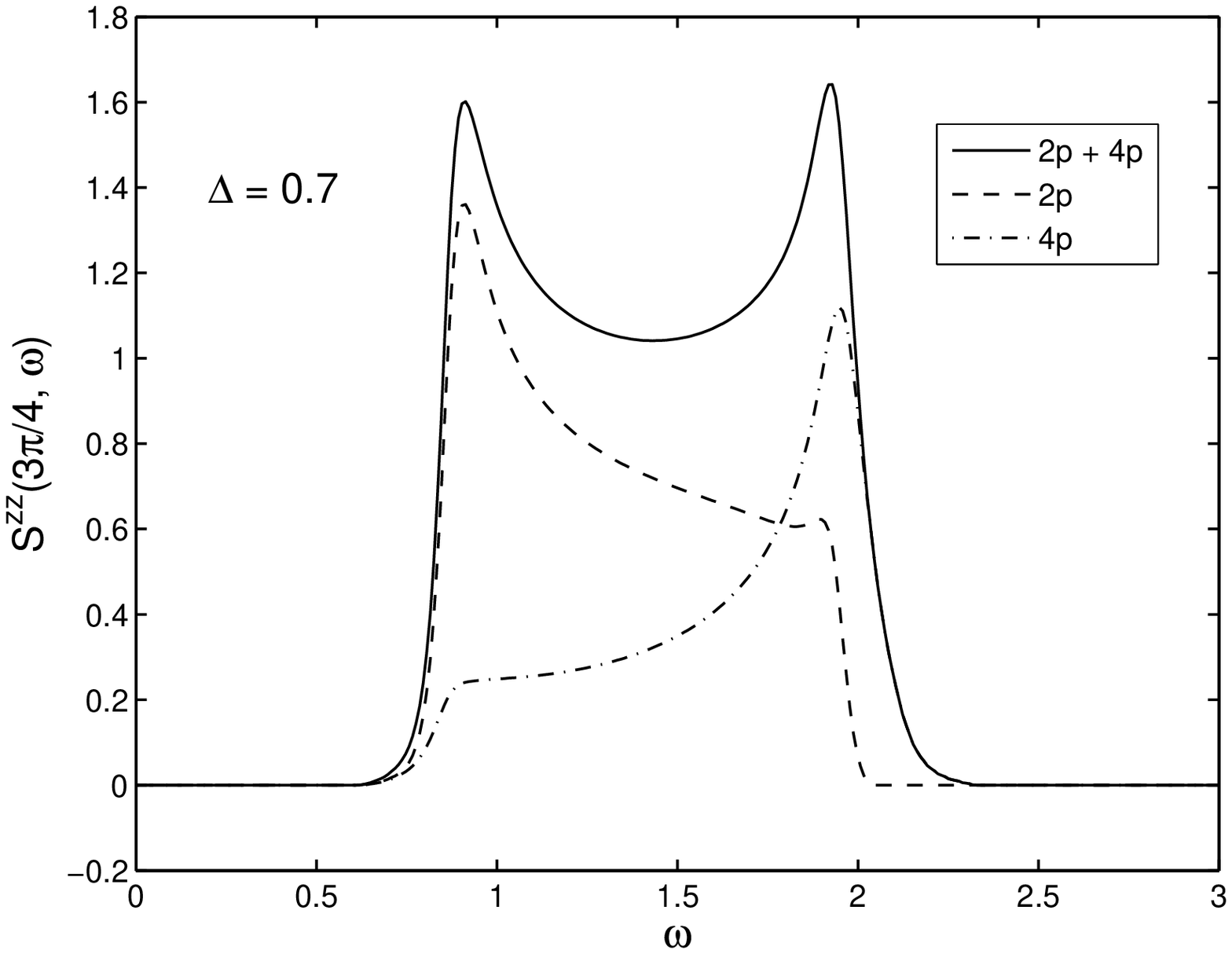}
\end{tabular}
\caption{Constant momentum $q$ slices of the structure factor as a function of frequency $\omega$, at $q = \pi/2$ 
and $3\pi/4$, for $XXZ$ with $\Delta = 0.3$ and $\Delta = 0.7$ at $M = N/4$.  2p denotes two-particle contributions, 
while 4p denotes four-particle contributions.  Resulting
curves are extrapolations to infinite size from $N = 4096$ (2p) and $384$ (2p + 4p).  2p contributions carry most of
the weight at $q = \pi/2$, but 4p ones become comparable at $q = 3\pi/4$.}
\label{Fixedq}
\end{figure*}

\begin{figure*}
\begin{tabular}{ll}
\includegraphics[width=8cm]{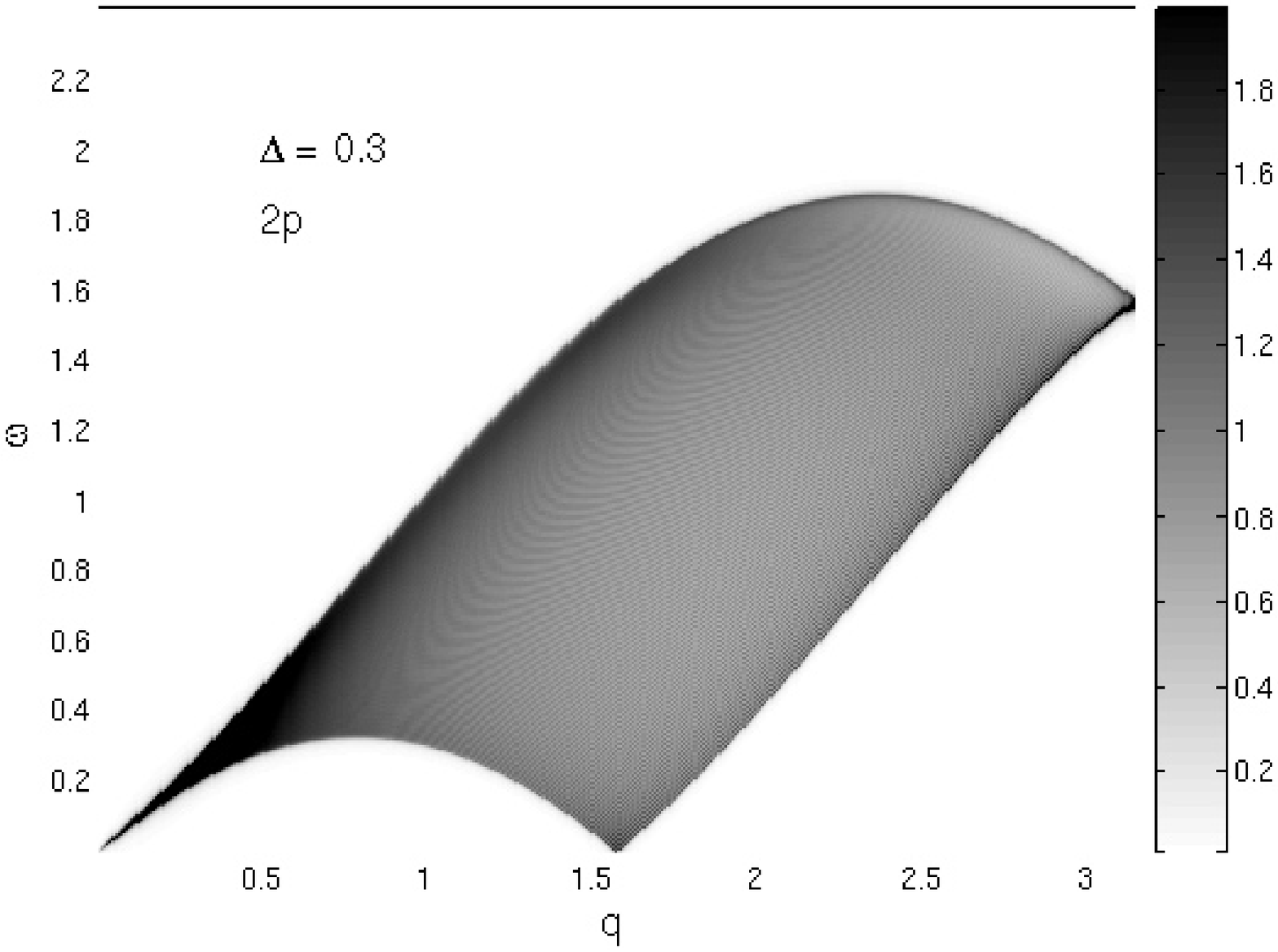}
& \includegraphics[width=8cm]{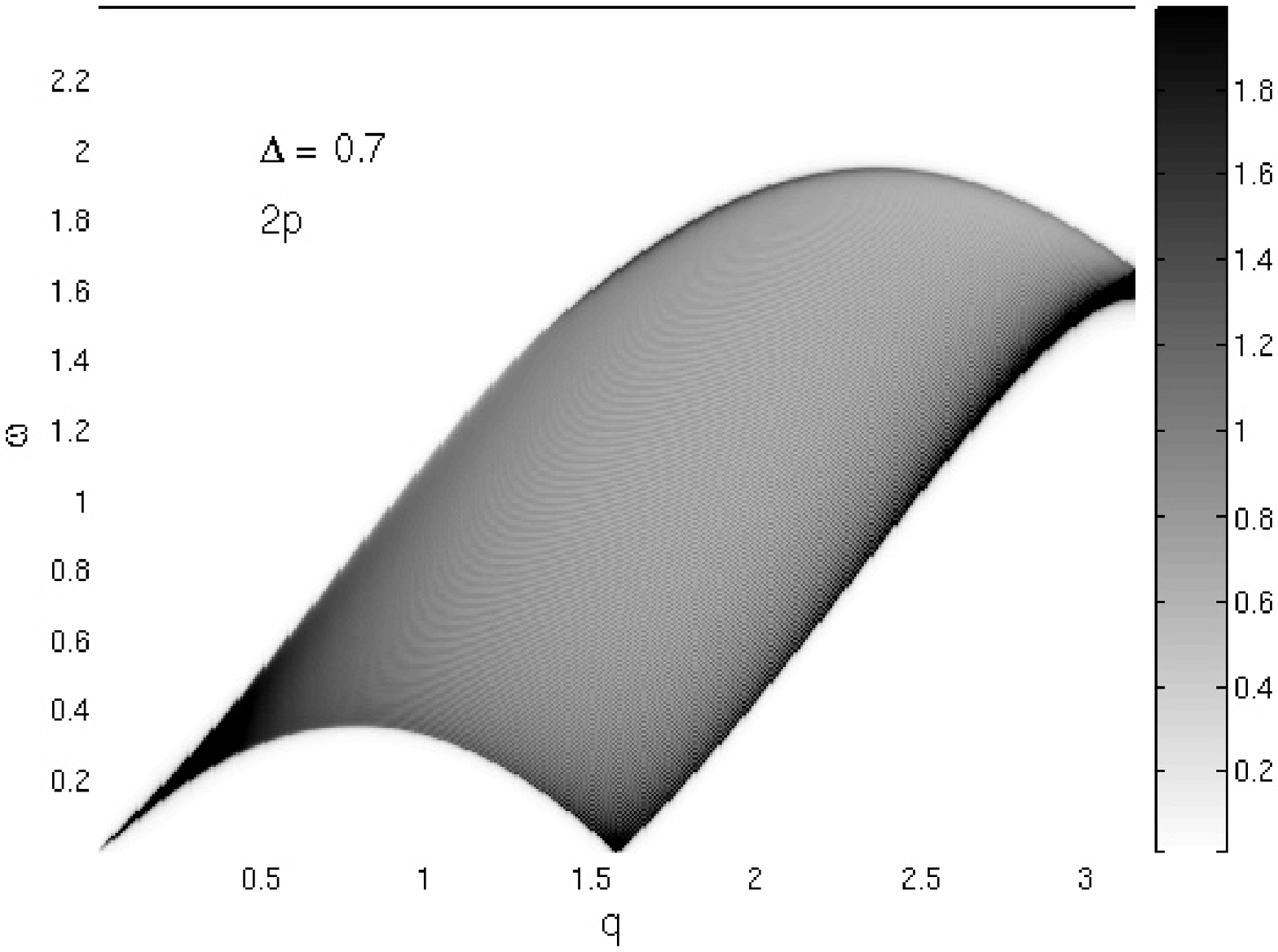} \\
\includegraphics[width=8cm]{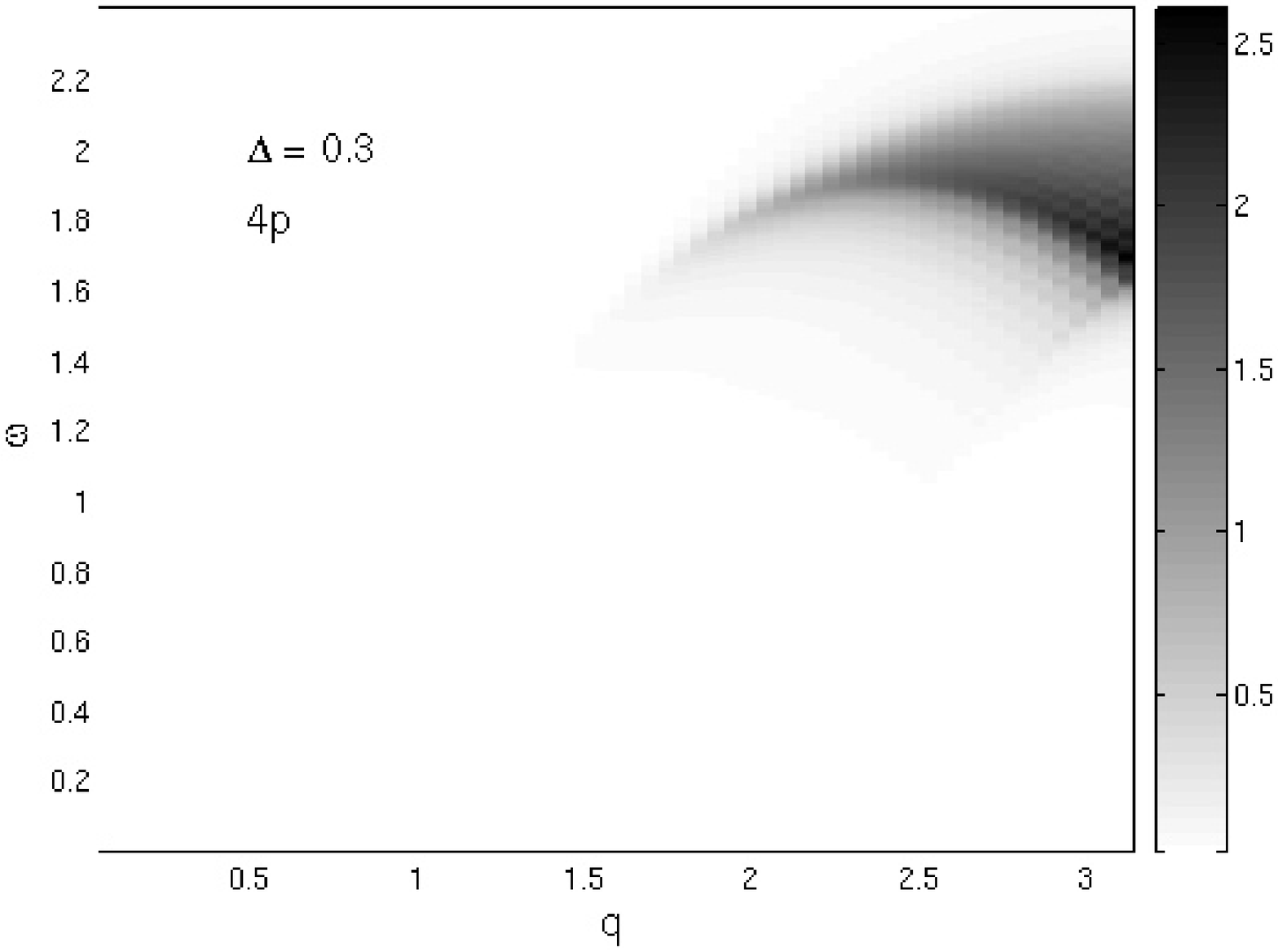} 
& \includegraphics[width=8cm]{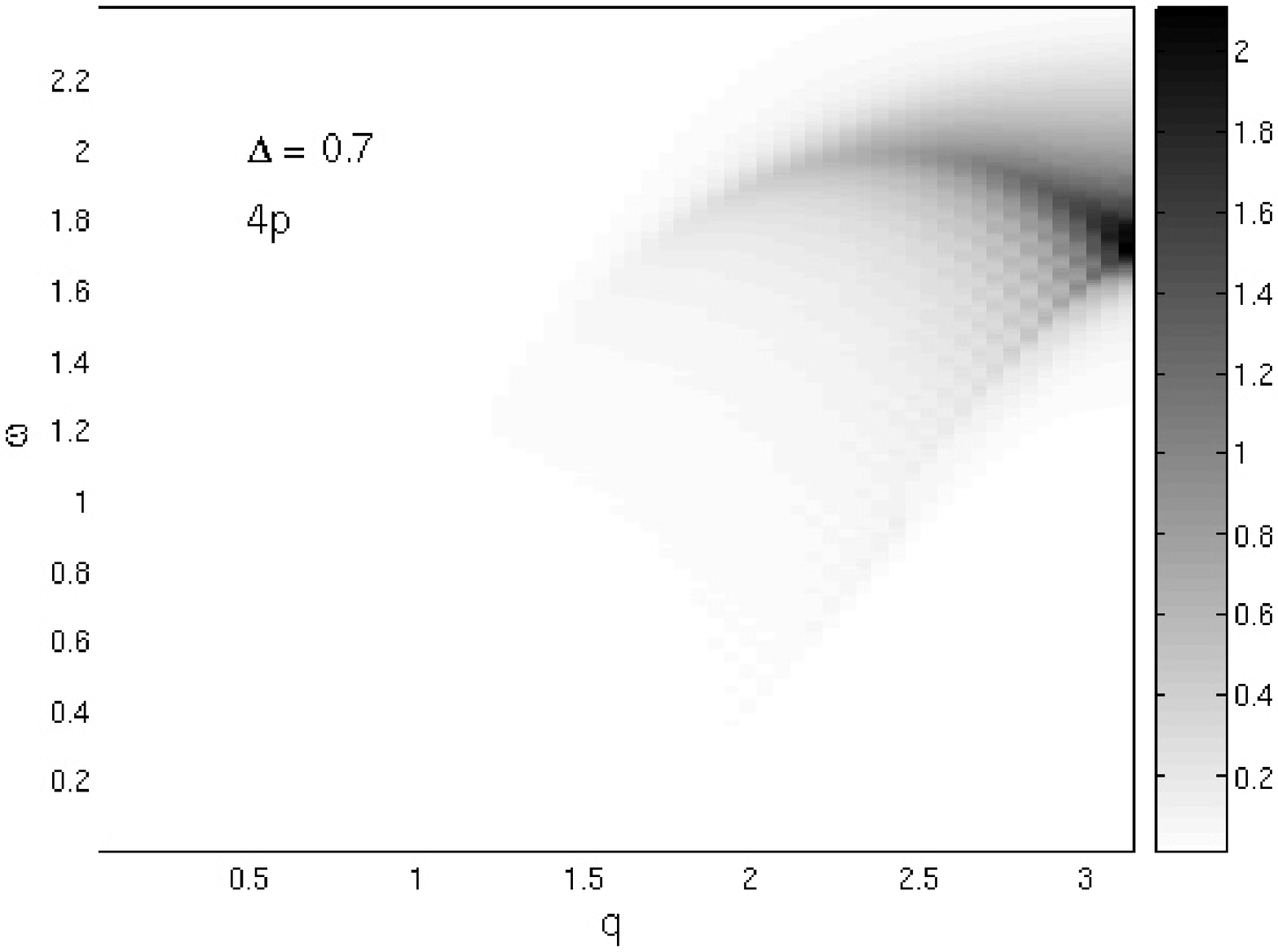} \\
\includegraphics[width=8cm]{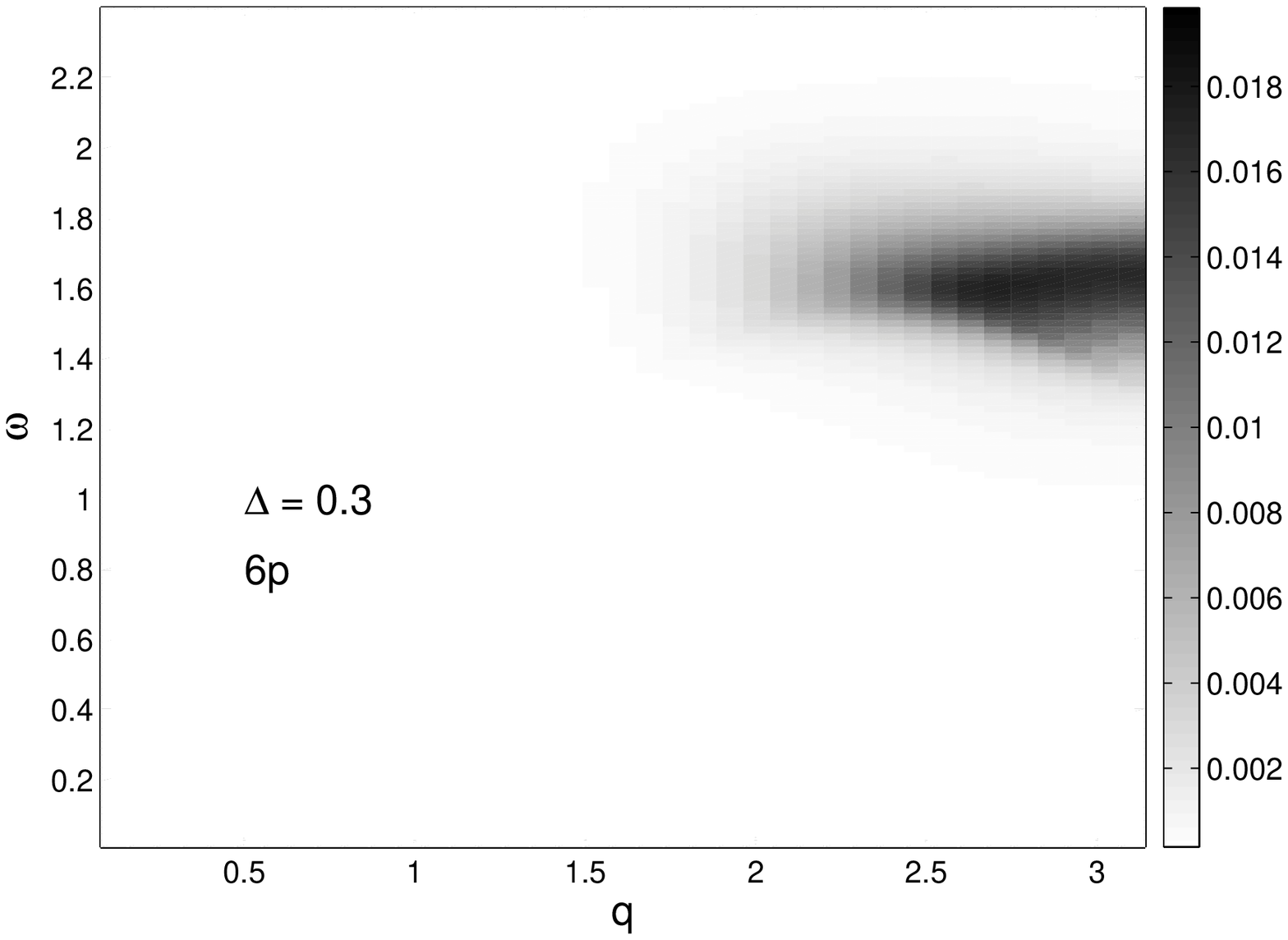} 
& \includegraphics[width=8cm]{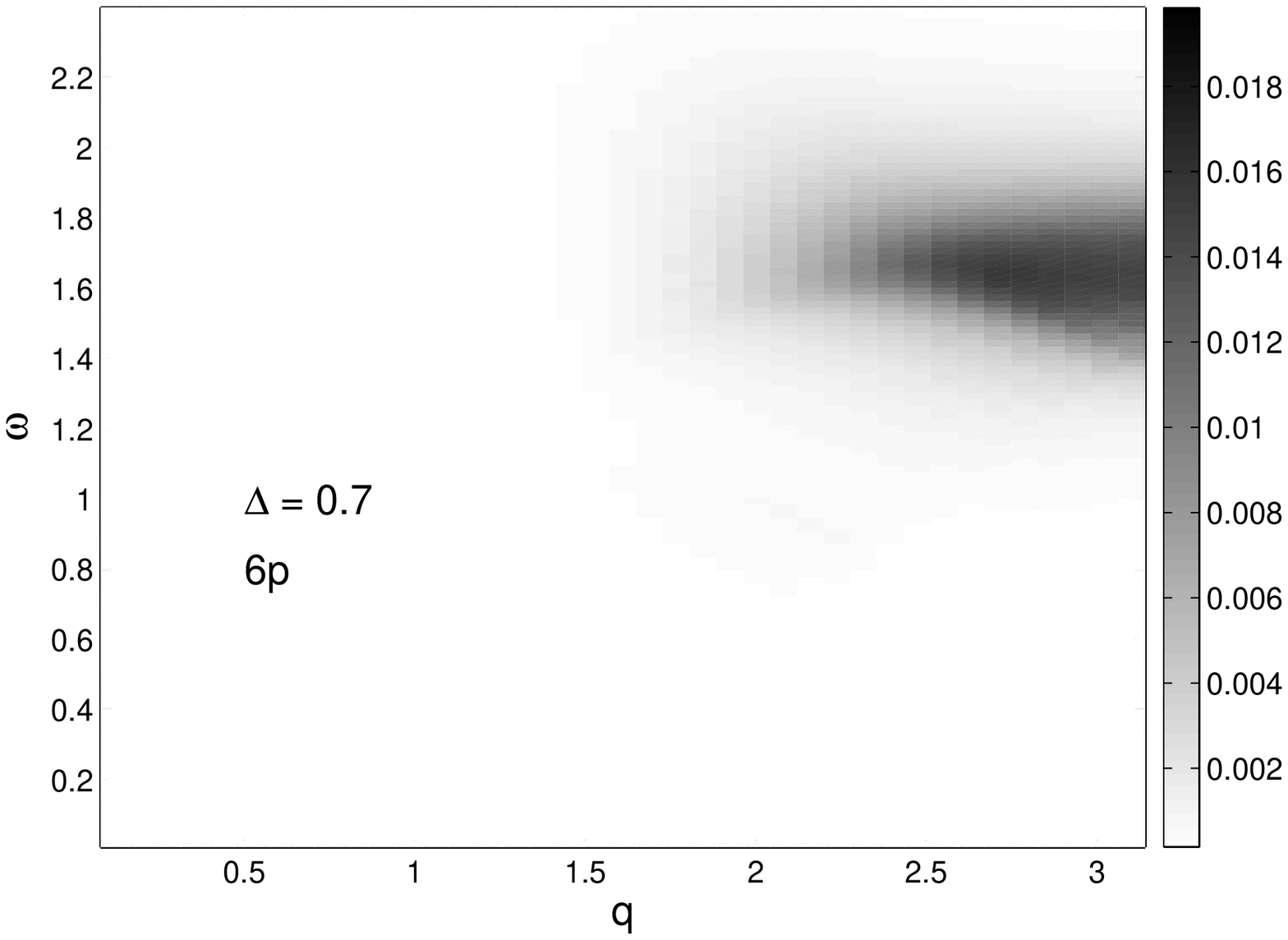} 
\end{tabular}
\caption{Longitudinal structure factor as a function of momentum $q$ and frequency $\omega$, for $XXZ$ with $\Delta = 0.3$, and
$\Delta = 0.7$ at $M = N/4$.  Two-, four- and six-particle contributions (labeled 2p, 4p and 6p) are plotted separately,
for system sizes $N = 768, 128$ and $80$ respectively.}
\label{Allq}
\end{figure*}

Obtaining the spin structure factor thus involves three steps:  scanning through
the eigenstates, solving the Bethe equations, and computing the determinants to obtain the form factor.
The fact that actually permits one to achieve a high degree of precision, is that the
contributions are rapidly decreasing functions of the number of particles involved.  
Contributions from unbound states to the structure factor come from even numbers of one-particle excitations.  
Two-particle contributions are those for which there is only one hole in the distribution of $I$'s in the
interval corresponding to the ground state.  
Similarly, four- and six-particle contributions are those with two and three holes
in the distribution of the $I$'s.  
It is well-known that the Bethe equations (\ref{BAE_log}) yield, for certain choices of $I$'s, complex rapidities 
in the form of strings (bound states).  This leads to reductions in the determinants (we will publish
the formulas elsewhere).  In the presence of a high field, the bound-state contributions are strongly suppressed.  
We compute the two-string contributions to show that this is indeed the case.  The zero-field case, where bound
states carry more weight, will be studied in a separate publication.

We present results for two different cases:  a chain with $\Delta = 0.3$, and another 
with $\Delta = 0.7$, both in a field such that $M = N/4$.  
For each case, we give two slices of the structure factor at fixed momentum $q = \pi/2, 3\pi/4$
as a function of frequency in Fig. \ref{Fixedq}, including contributions from up to four particles (six-particle
and bound-state contributions are too small).  We subsequently give density plots for the whole region $q \in [0, \pi]$
in Fig. \ref{Allq} for the same cases, including up to six-particle contributions.  In all cases, we use eq. (\ref{S_zz_ff}) where
we broaden the delta functions to width $\log N/N$ in order to obtain smooth curves.  The isotropic case $\Delta = 1$
again requires special treatment, which we will present elsewhere.  

Computing the structure factor for all momenta allows us to check the saturation of the sum rule
\begin{eqnarray}
\int_{-\infty}^{\infty} \frac{d\omega}{2\pi} \frac{1}{N}\sum_{q} S^{zz}(q, \omega) 
= \frac{1}{4}\left[1 - (1-\frac{2M}{N})^2\right] \equiv C.
\label{sumrule}
\end{eqnarray}
For $M = N/4$, the sum rule requires $C = 0.1875$.
\begin{table}
\caption{Contributions of two-, four- and six-particle sectors to the sum rule for $\Delta = 0.3$
(\%).}
\begin{tabular}{|l|l|l|l|l|}
\hline
N & 2p & 4p & 6p & Total \\ \hline
768 & 70.29 & & & 70.3 \\ \hline
128 & 78.86 & 20.25 & & 99.11 \\ \hline
80 & 82.17 & 17.52 & 0.23 & 99.92 \\ \hline
\end{tabular}
\label{SumRuleTable_D_0.3}
\end{table}
\begin{table}
\caption{Contributions of two-, four- and six-particle sectors to the sum rule for $\Delta = 0.7$
(\%).}
\begin{tabular}{|l|l|l|l|l|}
\hline
N & 2p & 4p & 6p  & Total \\ \hline
768 & 69.85 & & & 69.85 \\ \hline
128 & 81.43 & 17.17 & & 98.60 \\ \hline
80 & 85.19 & 14.02 & 0.28 & 99.49 \\ \hline
\end{tabular}
\label{SumRuleTable_D_0.7}
\end{table}

We compute two-particle contributions to the structure factor for system sizes up to $N = 768$.
Four- and six-particle ones are computed up to $N = 128$ and $80$ respectively.
Two-particle contributions are monotonically 
decreasing with system size, whereas four- and six-particle ones first increase, then decrease, 
due to the interplay between the finite-size gap and state counting.
The contribution to the sum rule from states with one bound state and two holes is computed
up to $N = 256$, and yields only $4.4 \times 10^{-7}$\% for $\Delta = 0.3$, and $3.0 \times 10^{-7}$\% for $\Delta = 0.7$.
Relative contributions are presented in Tables \ref{SumRuleTable_D_0.3}, \ref{SumRuleTable_D_0.7}.  
Two-particle terms account for about 70\% of the sum rule,
close to the 72.89\% result for the isotropic chain in zero field \cite{BougourziPRB54,KarbachPRB55}.
Including four-particle contributions saturates around 99\% of the sum rule for 128 sites.
Our results for 80 sites, including up to six-particle contributions, saturate well over 99\% of the total sum rule. 

In summary, we have numerically obtained correlation functions of integrable models on
the lattice using determinant representations for form factors, in the fundamental case of the anisotropic Heisenberg chain
in a field.  We have shown that by including multiparticle contributions, 
the sum rule can be saturated to very high precision.

\begin{acknowledgments}
J.-S. C. acknowledges support from CNRS and Stichting voor Fundamenteel Onderzoek der Materie (FOM).
J.-M. M. is a member of CNRS (UMR 5672), and acknowledges support from the EUCLID EC network.
\end{acknowledgments}


\begin{thebibliography}{99}
\bibitem{BetheZP71} H. Bethe, Z. Phys. 71, 205 (1931).
\bibitem{KorepinBOOK} see {\it e.g.} V. E. Korepin, N. M. Bogoliubov and A. G. Izergin, ``Quantum Inverse Scattering
Method and Correlation Functions'', Cambridge University Press, 1993; 
M. Takahashi, ``Thermodynamics of One-Dimensional Solvable Models'', Cambridge
University Press, 1999, and references therein.
\bibitem{MattisBOOK} D. C. Mattis, ``The Many-Body Problem'', World Scientific (1993).
\bibitem{EsslerREVIEW} F. H. L. Essler and R. M. Konik, in ``Ian Kogan Memorial Volume'', World Scientific (2005); cond-mat/0412421.
\bibitem{JimboBOOK} M. Jimbo and T. Miwa, ``Algebraic Analysis of Solvable Lattice Models'', AMS, Providence (1995).
\bibitem{MailletNPB554} N. Kitanine, J.-M. Maillet and V. Terras, Nucl. Phys. B 554, 647 (1999).
\bibitem{MailletNPB567} N. Kitanine, J.-M. Maillet and V. Terras, Nucl. Phys. B 567, 554 (2000).
\bibitem{KorepinCMP86} V. E. Korepin, Commun. Math. Phys. 86, 391 (1982).
\bibitem{KenzelmannPRB65} M. Kenzelmann, R. Coldea, D. A. Tennant, D. Visser, M. Hofmann, P. Smeibidl
and Z. Tylczynski, Phys. Rev. {\bf B} 65, 144432 (2002).
\bibitem{StonePRL91} M. B. Stone, D. H. Reich, C. Broholm, K. Lefmann, C. Rischel, C. P. Landee and M. M. Turnbull,
Phys. Rev. Lett. 91, 037205 (2003).
\bibitem{BiegelEPL59} D. Biegel, M. Karbach and G. M\"uller, Europhys. Lett. 59, 882 (2002).
\bibitem{BiegelJPA36} D. Biegel, M. Karbach and G. M\"uller, J. Phys. A: Math. Gen. 36, 5361 (2003).
\bibitem{SatoJPSJ73} J. Sato, M. Shiroishi and M. Takahashi, J. Phys. Soc. Jpn 73, 3008 (2004).
\bibitem{BougourziPRB54} A. H. Bougourzi, M. Couture and M. Kacir, Phys. Rev. B 54, R12669 (1996).
\bibitem{KarbachPRB55} M. Karbach, G. M\"uller, A. H. Bougourzi, A. Fledderjohann and K.-H. M\"utter, Phys. Rev. B 55, 12510 (1997).
\end{thebibliography}
\end{document}